\PassOptionsToPackage{numbers,sort&compress}{natbib}
\documentclass[final,5p,times,twocolumn]{elsarticle}
\journal{Physics Letters B}
\usepackage{amsmath}
\usepackage{amssymb}
\usepackage{graphicx}
\usepackage[linkcolor=blue,
            citecolor=blue,
            urlcolor=blue,
            colorlinks=true,
            breaklinks
            ]{hyperref}
\usepackage{slashed}
\usepackage{braket}

\def\1eq#1{(\ref{#1})}
\def\1eq#1{Eq.\nobreak\thinspace(\ref{#1})}

\def\2eqs#1#2{Eqs.\nobreak\thinspace(\ref{#1}) and\nobreak\thinspace(\ref{#2})}
\def\3eqs#1#2#3{Eqs.\nobreak\thinspace(\ref{#1}),\nobreak\thinspace(\ref{#2}) and\nobreak\thinspace(\ref{#3})}
\def\eq#1{Eq.\nobreak\thinspace(\ref{#1})}

\newcommand{\be}{\begin{equation}}
	\newcommand{\bea}{\begin{eqnarray}}
		\newcommand{\ee}{\end{equation}}
	\newcommand{\eea}{\end{eqnarray}}
    \def\s#1{{\scriptscriptstyle #1}}

\def\eg{{\it e.g.}, }
\def\fig#1{\hyperref[#1]{Fig.\nobreak\thinspace\ref*{#1}}}
\def\figs2#1#2{\hyperref[#1]{Figs.\nobreak\thinspace\ref*{#1}}\nobreak\thinspace and\nobreak\thinspace\hyperref[#2]{\ref*{#2}}}
\def\tab#1{\hyperref[#1]{Tab.\nobreak\thinspace\ref*{#1}}}
\def\sect#1{\hyperref[#1]{Sec.\nobreak\thinspace\ref*{#1}}}
\def\appref#1{\hyperref[#1]{App.\nobreak\thinspace\ref*{#1}}}

\providecommand{\Eq}[1]{\hyperref[#1]{Eq.~(\ref{#1})}}
\providecommand{\Int}{\mathcal{I}}

\providecommand{\aT}{\tilde{\alpha}_T}
\begin{document}
%%%%%%%%%%%%%%%%%%%%%%%%%%%%%%%%%%%%%%%%%%%%%%%%%%%%
% Title & Author list
%%%%%%%%%%%%%%%%%%%%%%%%%%%%%%%%%%%%%%%%%%%%%%%%%%%%
\begin{frontmatter}
\title{Light baryon masses from the modified Taylor coupling}
\author[uv]{A.S.~Miramontes}
\ead{angel.s.miramontes@uv.es}

\author[uv]{J.~Papavassiliou}
\ead{joannis.papavassiliou@uv.es}

\author[sal]{H.~Sanchis-Alepuz}

\affiliation[uv]{
  organization={\mbox{Department of Theoretical Physics and IFIC, University of Valencia and CSIC}},
  city={Valencia},
  postcode={E-46100},
  country={Spain}
}

\affiliation[sal]{
  organization={Silicon Austria Labs GmbH},
  addressline={Sandgasse 34},
  city={Graz},
  postcode={8010},
  country={Austria}
}
%%%%%%%%%%%%%%%%%%%%%%%%%%%%%%%%%%%%%%%%%%%%%%%%%%%%
% Abstract
%%%%%%%%%%%%%%%%%%%%%%%%%%%%%%%%%%%%%%%%%%%%%%%%%%%%
\begin{abstract}
We present a calculation of the positive-parity baryon octet and 
decuplet spectra, using a 
non-perturbative interaction governed by the modified Taylor 
effective charge and complemented by explicit flavour-dependent contributions. 
The framework transitions seamlessly from the mesonic domain, where it was first 
established, to the baryonic sector, requiring no further parameter optimization. 
By solving the relativistic three-body Faddeev equation, we extract both the 
ground-state masses and amplitudes for the corresponding baryonic 
configurations. The resulting spectrum reproduces experimental data within a few 
percent. Additionally, we derive the evolution of the nucleon and $\Delta$ masses 
as functions of the pion mass, contrast our findings directly against 
lattice-QCD benchmarks, and evaluate the propagation of the lattice 
uncertainties into the computed hadronic observables.

\end{abstract}

\end{frontmatter}
%%%%%%%%%%%%%%%%%%%%%%%%%%%%%%%%%%%%%%%%%%%%%%%%%%%%
% Introduction
%%%%%%%%%%%%%%%%%%%%%%%%%%%%%%%%%%%%%%%%%%%%%%%%%%%%
\section{Introduction}\label{sec:intro}

The positive-parity baryon octet ($J^P = \tfrac{1}{2}^+$) and decuplet
($J^P = \tfrac{3}{2}^+$) comprise eighteen states whose masses are
experimentally well-established \cite{ParticleDataGroup:2026mpi}, 
providing an essential benchmark for
validating non-perturbative approaches to the QCD dynamics. 
Within lattice QCD, 
the baryonic spectrum has been
studied extensively, and the
light masses are accurately reproduced ~\cite{Durr:2008zz,Fodor:2012gf,Edwards:2011jj,BMW:2014pzb}. Functional methods \cite{Huber:2025cbd,Eichmann:2025wgs}
offer a complementary continuum 
perspective 
by constructing hadronic observables from appropriate combinations of the 
Green's (correlation) 
functions of the theory.
In this framework, 
the key Green's functions 
are determined via 
suitable truncations of the 
Schwinger-Dyson equations (SDEs)  
that govern their 
evolution~ \cite{Alkofer:2000wg,Fischer:2006ub,Williams:2015cvx,
Eichmann:2016yit, Huber:2018ned,
Miramontes:2025imd,Ferreira:2026gbe,Wessely:2026cqq}.  

Within standard implementations of the functional framework, 
the dressed quark propagator is computed from the so-called 
gap equation, while bound states
appear as poles in the corresponding Green's functions. These bound states are 
described by the Bethe-Salpeter equation (BSE) for mesons
~\cite{Salpeter:1951sz,PhysRev.84.350,Bethe1957,Nakanishi:1969ph,Jain:1993qh,Munczek:1994zz,Alkofer:2002bp,Krassnigg:2009zh,Fischer:2014xha,Miramontes:2022mex,Eichmann:2025wgs, Huber:2025cbd,Hagel:2025ngi}
and by the relativistic Faddeev equation for baryons~
\cite{Faddeev:1960su,Eichmann:2009qa,Sanchis-Alepuz:2011egq,Sanchis-Alepuz:2013iia,Sanchis-Alepuz:2014wea,Sanchis-Alepuz:2014sca,Eichmann:2016hgl,Yao:2024ixu}.
From an operational point of view, the baryonic case is considerably more demanding, 
since it involves three valence quarks and possesses an extended tensorial structure; 
consequently, comparatively few studies have systematically addressed the complete 
set of octet and decuplet states
~\cite{Sanchis-Alepuz:2014sca,Sanchis-Alepuz:2015fcg,Eichmann:2016yit}.
This three-body problem may be solved directly, with the three quarks treated symmetrically, or reduced to an effective two-body problem through the
quark-diquark approximation, which assumes that correlations within quark pairs
dominate \cite{Oettel:1998bk,Eichmann:2016yit,Eichmann:2016hgl,Gutierrez-Guerrero:2019uwa,Liu:2023reo}. 
Regardless of the chosen strategy, however, both frameworks rely crucially on the 
same fundamental prerequisite: the reliable modeling of the 
non-perturbative 
interaction between quarks and gluons.

Historically, the octet and the decuplet baryon masses have been studied 
predominantly within the rainbow-ladder
(RL) approximation, where  
the full quark-gluon vertex,
$\Gamma_{\mu} (q,r,p)$,
is drastically reduced to its tree-level expression,  $\Gamma_{\mu} (q,r,p) \to \gamma_{\mu}$.
Within this framework,
the fundamental interaction 
between quarks and gluons is modeled by a 
phenomenological kernel, whose 
infrared parameters are calibrated to reproduce 
ground-state hadronic observables
 (\eg masses 
and decay constants) 
of  the meson sector \cite{Maris:1997tm, Maris:1999bh, Maris:1999nt}.
Incorporated into 
the Faddeev equation, this RL kernel yields baryon masses that differ from 
the experimental values by a few percent, manifesting a maximum discrepancy of 
approximately $10\%$ in the $\Sigma$ state~\cite{Sanchis-Alepuz:2014sca}. 

In the ongoing quest for a fundamental description, considerable effort has been 
dedicated to the systematic inclusion of QCD-derived information into the 
kernels governing the formation of hadronic bound states~\cite{Heupel:2014ina,Sanchis-Alepuz:2015tha,Ferreira:2026gbe,Ferreira:2026zpr,Miramontes:2025ofw,Wessely:2026cqq}. 
This is accomplished by determining the non-perturbative structure of the 
quark-gluon vertex, $\Gamma_{\mu}(q,r,p)$, via its corresponding SDE, the 
Slavnov-Taylor identities (STIs), or a combination thereof. Crucially, this 
information must be incorporated into the bound-state equations in a 
symmetry-preserving manner, respecting the underlying chiral symmetry 
and its associated axial Ward-Takahashi identities (WTIs).

While the implementation of more elaborate vertex schemes is currently underway, 
considerable physical insight can be extracted from the simplified truncation 
of the quark-gluon vertex put forth in the recent work of~\cite{Gao:2024gdj}. 
Specifically, the approximated $\Gamma_{\mu}(q,r,p)$ retains exclusively the 
dominant form factor  associated with the tree-level tensor 
structure $\gamma_{\mu}$. In the symmetric kinematic limit, this form factor is determined 
from the corresponding STI, and subsequently incorporated into the standard 
one-gluon exchange kernel of the meson BSEs. The resulting interaction is 
composed of two distinct elements: $(\it{i})$ a modified version of the 
standard Taylor coupling~\cite{vonSmekal:1997ohs,Boucaud:2008gn,vonSmekal:2009ae,Blossier:2012ef,Zafeiropoulos:2019flq}, 
denoted by $\tilde{\alpha}_{\s T}(q^2)$, and $(\it{ii})$ explicit 
\textit{flavour-dependent} contributions expressed in terms of the  
quark dressing functions. The implementation of this framework in the meson 
sector successfully reproduces the masses, decay constants, and electromagnetic 
form factors of heavy-light systems~\cite{Gao:2024gdj,Miramontes:2025vzb,Miramontes:2026san}.

In this work, we generalize this framework to the baryonic sector by 
incorporating the flavour-dependent interaction directly into the  
three-body Faddeev equation. Specifically, we compute the masses 
of the positive-parity octet and decuplet baryons, spanning all possible valence 
configurations of light and strange quarks. We emphasize that the transition 
from the mesonic to the baryonic sector takes place seamlessly, without any 
additional parameter tuning: the effective Taylor charge remains fixed by its 
lattice-QCD definition, while the current quark masses are preserved identically 
from the mesonic sector. The resulting spectrum agrees with experimental data 
at the level of a few percent.
In comparison to previous RL studies, the calculated masses move systematically,
 albeit slightly, closer to
the experimental values; 
the most pronounced improvement is manifested in the  strange members of the octet, where the explicit flavour dependence of the 
underlying interaction becomes particularly relevant.

In addition to the spectrum, 
we extract the corresponding baryonic amplitudes 
directly from the Faddeev equation.
These quantities encode the internal structure of the states in terms of the
momenta, spin and flavours of the valence quarks. They constitute the primary
input for evaluating the matrix elements of external currents, such as
electromagnetic, axial and transition form factors
\cite{Eichmann:2011vu,Sanchis-Alepuz:2015fcg,Sanchis-Alepuz:2017jjd,Sanchis-Alepuz:2017mir,Eichmann:2016yit,Miramontes:2021xgn,Chen:2023zhh,Xu:2024fun,Miramontes:2025ofw,Albino:2025fcp,Yao:2026phv}. For strange baryons, such observables represent a central goal of the
experimental program at FAIR, where the production and electromagnetic
decays of hyperons are actively investigated by the HADES collaboration
\cite{HADES:2020pcx,Lalik:2019yag}.

The article is organized as follows. \sect{sec:framework} summarizes the 
fundamental ingredients of the calculation, introduces the flavour-dependent 
interaction strength, and sets up the relativistic Faddeev equation. \sect{sec:results} details the numerical 
procedure and presents the calculated octet and decuplet spectra, which are 
contrasted directly against RL results and experimental data. 
Furthermore, we analyze the dominant covariants of the mixed-antisymmetric
nucleon Faddeev amplitude and compute the masses of the nucleon and the
$\Delta$ as functions of the pion mass, for direct comparison with lattice QCD. In addition, we briefly 
evaluate the propagation of uncertainties originating from the lattice determination 
of the Taylor coupling into our hadronic observables. Finally, our conclusions and 
an outlook on future extensions are presented in \sect{sec:outlook}.

\section{General framework}\label{sec:framework}

In this section we 
briefly overview the 
framework employed to compute the baryonic masses and corresponding bound-state 
amplitudes.
Specifically, we define the 
effective interaction strength, and outline the key
dynamical equations where it enters---namely, the quark gap equation and the 
relativistic three-body Faddeev equation.

\subsection{Interaction strength and modified Taylor coupling} \label{sec:interaction}

\vspace{0.05cm}

According to the original formulation of \cite{Gao:2024gdj}, 
after factoring out the standard color structure, the quark-gluon vertex 
is approximated by  
\be
{\Gamma}_{\!f}^{\,\mu}(q,r,p) 
= \lambda_{f}(q,r,p) \gamma^{\mu} + \cdots\,, 
\label{eq:classic}
\ee
where the ellipsis denotes all 
remaining tensorial components,
and the index $f=u,d,s,c,b$ stands for the 
flavour of the quark involved. 
The form factor 
$\lambda_{f}(q,r,p)$ is subsequently evaluated in the so-called ``symmetric configuration'', 
defined by \mbox{$q^2=r^2=p^2$}. 
Under certain simplifying assumptions,
the STI satisfied by
the quark-gluon vertex yields the approximate expression 
\begin{align}
{\lambda}^{\mathrm{sym} }_f(q^2)
=  F(q^2) A_{f}(q^2) 
{\tilde X}_0(q^2) \,, 
\label{eq:gtlsym}
\end{align}
where $F(q^2)$ is the dressing function of the ghost field in the Landau gauge,  and  
$A_{f}(q^2)$ is the quark dressing function, see \1eq{eq:quarkprop}.
The function ${\tilde X}_0(q^2)$ represents 
the flavour-independent part of the 
ghost-gluon kernel, and its integral expression is given in Eq.~(14) of \cite{Gao:2024gdj}. 

In the one-gluon-exchange approximation, the gluon propagator is multiplied 
by the two quark-gluon vertices 
on its ends to furnish the 
total effective strength.
Since the Landau-gauge gluon propagator is given by   
\be
\Delta_{\mu\nu}(q) = {\cal Z}(q^2)\,\Delta^{(0)}_{\mu\nu}(q) \,,\qquad
\Delta^{(0)}_{\mu\nu}(q)=
\left( \delta_{\mu\nu} - \frac{q_\mu q_\nu}{q^2} \right) \frac{1}{q^2}\,,
\label{glueprop}
\ee
with ${\cal Z}(q^2)$ 
denoting the gluon dressing function,  
the scalar part of the resulting 
effective interaction, $\Int_{\!ff'}(q^2)$,
is given by 
\begin{equation}
  \Int_{\! ff'}(q^2) = {\tilde\alpha}_{\s T}(q^2)\, A_f(q^2)\, A_{f'}(q^2) \,.
\label{eq:interaction}
\end{equation}
The quantity ${\tilde\alpha}_{\s T}(q^2)$ 
represents the {\it modified} Taylor coupling, related to the standard $\alpha_{\s T}(q^2)$ by 
\be
{\tilde\alpha}_{\s T}(q^2)= 
\alpha_{\s T}(q^2){\tilde X}_0^2(q^2) \,,
\qquad
\alpha_{\s T}(q^2) =
\alpha_s\, {\cal Z}(q^2)\, F^2(q^2)
\,,
\label{eq:modTaylor}
\ee
where $\alpha_s = g^2/4 \pi$ 
denotes the strong coupling constant.
Note that the entire flavour dependence of $\Int_{\! ff'}(q^2)$
is contained in the quark dressing functions
$A_f(q^2)$ and $A_{f'}(q^2)$, whose structure is obtained dynamically by solving the gap equation.

%%%%%%%%%%%%%%%%%%%%%%%%%%%%%%%%%%%%%%%%%%%%%%%%%%%%
\begin{figure*}[t!]
  \centering
  \includegraphics[width=\textwidth]{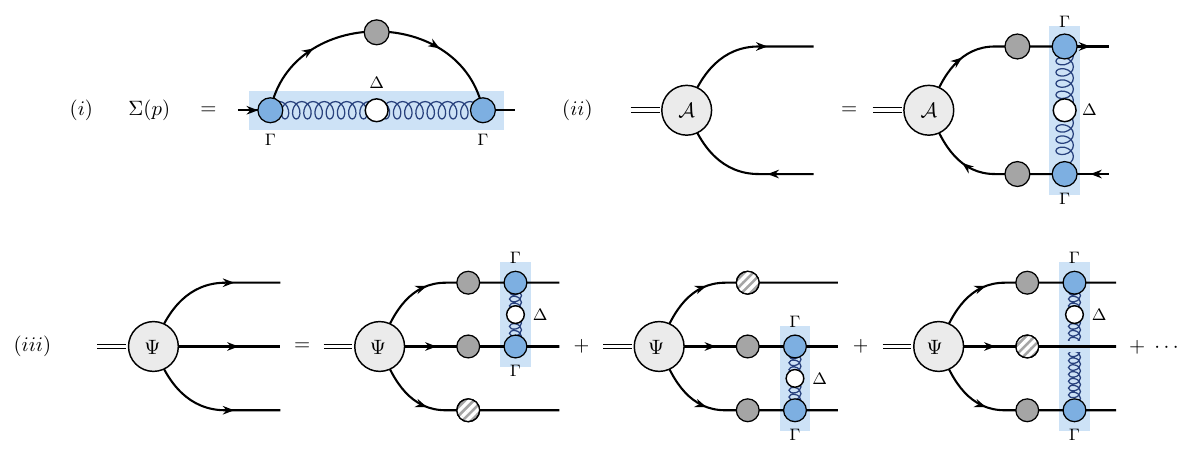}
 \caption{Diagrammatic representation of the relevant dynamical equations. Panel $(i)$
shows the quark gap equation, panel $(ii)$ the meson BSE,
and panel $(iii)$ the three-body Faddeev equation. The blue bands denote the
interaction kernel of \1eq{eq:kernel}, built out of the fully-dressed
quark-gluon vertices $\Gamma$ (blue circles) and the gluon propagator $\Delta$
(white circles). Grey circles stand for the dressed quark propagators, while 
hatched circles denote the spectator quark. The ellipsis in panel $(iii)$
indicates the irreducible three-body contributions contained in the second term of \1eq{eq:faddeev}, 
which are not included in
the present work.}
  \label{fig:equations}
\end{figure*}
%%%%%%%%%%%%%%%%%%%%%%%%%%%%%%%%%%%%%%%%%%%%%%%%%%%%

Restoring the tensorial content of the interaction, one obtains the  
interaction kernel 
\begin{equation}
  \mathcal{K}_{ff'}(q) =
  4\pi\, \gamma^\mu\, \Int_{ff'}(q^2)\, \gamma^\nu\,
  \Delta^{(0)}_{\mu\nu}(q) \,,
  \label{eq:kernel}
\end{equation}
which is a common ingredient of all dynamical equations employed in our analysis, 
indicated by the blue bands in \fig{fig:equations}.

It is instructive to compare \1eq{eq:kernel} with the kernel of the RL
approximation. In both cases the two vertices enter only through their
classical tensor structure, so that the momentum dependence 
of the kernel is
that of the gluon propagator. Consequently, in the chiral limit both schemes preserve the axial WTI, which
guarantees a massless pion. For unequal current
quark masses, the flavour dependence of $\Int_{ff'}$ from \1eq{eq:interaction} leads to violations of the
generalized WTI, proportional to $m_f - m_{f'}$. In the meson sector their effect has been estimated at the level of $0.2\%$
for particles composed of $u/d$ and $s$ quarks and below $0.5\%$ for the
heaviest systems \cite{Miramontes:2025vzb}. The two kernels thus have the same tensor structure and differ only in the
interaction strength. In previous RL studies of baryons, a single function of $q^2$ is used for all
flavours, with infrared parameters adjusted to meson observables. Here,
instead, $\Int_{\!ff'}(q^2)$ is constructed from correlation functions of QCD
and depends on the flavours attached to the two vertices through the
corresponding quark dressing functions.

\subsection{Quark propagator} \label{sec:quark}

The propagator for any of the quark flavours entering a given baryon is obtained from
the corresponding gap equation. 
Employing the kernel in  \1eq{eq:kernel}, this integral equation takes
the form [see panel $(i)$ in \fig{fig:equations}]
\begin{equation}
  S^{-1}_{\!f}(p) = Z_{2,f}\left( i\slashed{p} + m_f \right)
  + 4\pi C_F \int_k \gamma^\mu S_{\!f}(k+p) \gamma^\nu
  \Delta^{(0)}_{\mu\nu}(k)\, \Int_{\!ff}(k^2) \,,
  \label{eq:gap}
\end{equation}
where $C_F = 4/3$ is the Casimir eigenvalue of the fundamental representation,
$m_f$ the current quark mass of the flavour $f$ and $Z_{2,f}$ the
corresponding quark renormalization constant. In addition, 
we introduced the short-hand notation  
\be
\int_k := (2\pi)^{-4}\int d^4k 
\label{intsh}
\ee
for the integration of the virtual momenta, where the use of 
a symmetry-preserving regularization scheme is understood.  

Following standard notation, the inverse quark propagator is
decomposed as
\begin{equation}
  S^{-1}_{\!f}(p) = i\slashed{p}\, A_{f}(p^2) + B_f(p^2) \,,
  \label{eq:quarkprop}
\end{equation}
with $A_f(p^2)$ and $B_f(p^2)$ the dressings of the quark propagator. The renormalization-group invariant 
quark mass function, $\mathcal{M}_f(p^2)$, is given by 
$\mathcal{M}_f(p^2) = B_f(p^2)/A_f(p^2)$. 

\Eq{eq:gap} is solved separately for
each of the flavours required by the physical system under consideration, namely $u$,
$d$, $s$, $c$ or $b$. In the present work we operate in the isospin-symmetric
limit, where $m_u = m_d$, so that only the light and the strange propagators are
required.

%%%%%%%%%%%%%%%%%%%%%%%%%%%%%%%%%%%%%%%%%%%%%%%%%%%%
\begin{figure*}[t!]
  \centering
  \includegraphics[width=\textwidth]{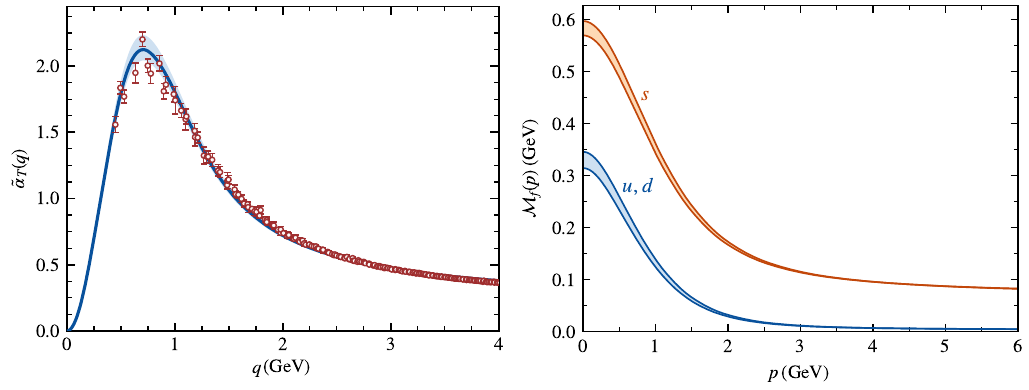}
\caption{Left: the modified Taylor effective charge, $\aT(q^2)$. The red points are the lattice $\alpha_{\s T}(q^2)$ \cite{Zafeiropoulos:2019flq}  multiplied by ${\tilde X}_0^2(q^2)$ (see \1eq{eq:modTaylor}), while the blue line is the parametrization
introduced in \cite{Gao:2024gdj}. Right: the mass functions
$\mathcal{M}_f(p)$ of the light and strange quarks obtained from
\1eq{eq:gap}, with $\mathcal{M}_u(0) = 0.34$~GeV and
$\mathcal{M}_s(0) = 0.59$~GeV, as in \cite{Gao:2024gdj}. In both panels, the
bands correspond to the variation of the fit parameters of $\aT$ by one
standard deviation around their central values.}
  \label{fig:alphaT}
\end{figure*}
%%%%%%%%%%%%%%%%%%%%%%%%%%%%%%%%%%%%%%%%%%%%%%%%%%%%

The renormalization of \1eq{eq:gap} is carried out by imposing at $p^2 = \mu^2$ the standard conditions
\begin{equation}
  A_f(\mu^2) = 1 \,, \qquad B_f(\mu^2) = m_f \,,
  \label{eq:renorm}
\end{equation}
which, in turn, provide the expressions for $Z_{2,f}$ and the mass
renormalization constant \cite{Eichmann:2025wgs}. 
The conditions in \1eq{eq:renorm} 
are part of the Taylor scheme~\cite{Taylor:1971ff, Marciano:1977su, Boucaud:2008gn},
employed in \cite{Gao:2024gdj}. As was done there, we fix the renormalization scale at $\mu = 4.3$ GeV. Most importantly, 
the current quark masses found in \cite{Gao:2024gdj} by fitting meson 
observables are employed in the present analysis,
{\it without} any further adjustments.

Let us finally point out that the bound state equations of \sect{sec:faddeev}
require knowledge of the quark propagators at complex values of their arguments, enclosed 
within parabolic domains that expand  as the mass of the bound state increases.
The corresponding analytic continuation is carried out by means of the Cauchy interpolation
method \cite{Sanchis-Alepuz:2017jjd,Fischer:2005en,Blank:2010bp}, whose range of
applicability is limited by the complex conjugate poles that $S_{\!f}$ develops in
the present truncation. These poles
appear to be closely connected to the approximation of the quark-gluon vertex solely by its classical
tensor structure. 
In general, 
vertex models of this type tend to exceed a critical
strength above which complex conjugate poles emerge \cite{Pawlowski:2024kxc}.
Instead, when the gap equation is solved in conjunction with the full quark-gluon vertex,
the complex conjugate poles are replaced by a pair of real poles with opposite-sign residues \cite{Wieland:2026iml,Alkofer:2026zhf}.
Clearly, this important point deserves 
further detailed analysis, to be 
presented elsewhere. 

\subsection{Bound-state equations} \label{sec:faddeev}

The interaction kernel shown in \1eq{eq:kernel} is employed for bound states of both mesons and
baryons. For a meson composed of a quark of flavour $f$ and an antiquark
of flavour $f'$, the Bethe-Salpeter amplitude $\mathcal{A}_{ff'}$ is determined from the
BSE displayed in panel $(ii)$ of \fig{fig:equations},
\begin{equation}
  \begin{split}
    \mathcal{A}_{ff'}(p_1,p_2) = -4\pi \int_k
    &\gamma^\mu S_{\!f}(k_1)\, \mathcal{A}_{ff'}(k_1,k_2)\,
     S_{\!f'}(k_2)\, \gamma^\nu \\
    &\times \Delta^{(0)}_{\mu\nu}(k)\, \Int_{\!ff'}(k^2) \,,
  \end{split}
  \label{eq:bse}
\end{equation}
with $k_i = k + p_i$. 

In the baryonic sector, bound states appear as poles in the six-point quark 
Green's function.
The corresponding bound-state amplitude, $\Psi$, satisfies 
the relativistic three-body equation
\begin{equation}
  \Psi = 
        \sum_{a=1}^{3}\mathcal{C}\,\mathcal{K}^{(2)}_{f_b f_c}\,G_0^{(2,a)}\,\Psi +\mathcal{K}^{(3)}\,G_0^{(3)}\,\Psi\,,
  \label{eq:faddeev}
\end{equation}
which is displayed schematically in panel $(iii)$ of \fig{fig:equations}.

The first term on the r.h.s. 
of \1eq{eq:faddeev} describes the 
part of the interaction mediated by 
two-body dynamics, where index $a$ runs over the three permutations 
of a spectator quark, while the remaining pair $(b,c)$ undergoes two-body 
interactions governed by the kernel of \eq{eq:kernel}.
 The factor $G_0^{(2,a)}$ denotes the product of the 
propagators, while $f_b$, $f_c$ indicate their flavours. The color factor
$\mathcal{C}$ is obtained by contracting the color singlet of the amplitude
with the color structure of the kernel.

The second term on the r.h.s. of \1eq{eq:faddeev}
captures the genuine three-body 
part of the interaction
\cite{Sanchis-Alepuz:2011egq, Eichmann:2009qa}; its 
diagrammatic representation is
absorbed in the 
ellipses of the panel $(iii)$ of \fig{fig:equations}. 
Specifically, $\mathcal{K}^{(3)}$ 
indicates the 
three-particle irreducible kernel, while  
$G_0^{(3)}$ denotes the product of the three dressed quark propagators.
Our knowledge of 
$\mathcal{K}^{(3)}$ is 
rather limited: the 
term containing the three-gluon vertex coupled to each of the quark lines is identically zero due to color traces,
while the rest is expected to be 
numerically subleading relative to the two-body contributions
\cite{Eichmann:2009qa,Sanchis-Alepuz:2011egq}. 

Throughout this work we adopt 
the so-called ``Faddeev approximation'',
where the three-particle kernel 
is assumed to vanish, $\mathcal{K}^{(3)}=0$.
The effective non-perturbative interaction
is then described exclusively by the two-body kernel of \1eq{eq:kernel}, which is attached to 
the three distinct permutations 
of interacting quark pairs.  

The bound-state amplitude $\Psi$ depends on the individual quark momenta $p_1$, $p_2$, and $p_3$.  These are traded for the total baryon momentum \mbox{$P=p_1+p_2+p_3$} and
two relative Jacobi momenta,
\begin{equation}
  p = (1-\zeta)\,p_3 - \zeta\,(p_1+p_2) \,, \qquad t = \frac{p_2-p_1}{2} \,,
  \label{eq:jacobi}
\end{equation}
with the momentum-partitioning parameter fixed to the symmetric value
$\zeta=1/3$. The total amplitude factorizes into colour, flavour and spin-momentum components. For the baryon octet it decomposes as 
\begin{equation}
  \Psi(p,t,P) = \Big[\, \Psi_{\rm MA}(p,t,P)\otimes F_{\rm MA}
  + \Psi_{\rm MS}(p,t,P)\otimes F_{\rm MS} \Big]
  \otimes \frac{\epsilon_{rst}}{\sqrt{6}} \,,
  \label{eq:octet}
\end{equation}
where $\epsilon_{rst}/\sqrt{6}$ represents the antisymmetric colour singlet, while 
$F_{\rm MA}$ and $F_{\rm MS}$ denote the mixed-antisymmetric and
mixed-symmetric SU(3) flavour wave functions of the state, respectively
\cite{Sanchis-Alepuz:2014sca}. For the decuplet, the expansion 
simplifies to a single term featuring a totally symmetric flavour state. Each spin-momentum component is expanded in an orthonormal covariant basis $\tau_i(p,t,P)$,
\begin{equation}
  \Psi_\rho(p,t,P) = \sum_i f_{\rho}^{\,i}(p^2,t^2,\{z\})\, \tau_i(p,t,P) \,,
  \label{eq:basis}
\end{equation}
where $\rho = {\rm MA}, {\rm MS}$, 
comprising $64$ elements for $J=1/2$ and $128$ for
$J=3/2$ \cite{Eichmann:2009zx,Sanchis-Alepuz:2011egq}. 

The dressing functions
$f_{\rho}^{\,i}$ depend on $p^2$, $t^2$ and on three angular variables formed between
the momenta $p$, $t$ and $P$, collectively denoted by $\{z\}$. The angular
dependence of these functions is expanded in a series of Chebyshev polynomials \cite{Sanchis-Alepuz:2017jjd}. Furthermore, 
the elements $\tau_i$ are classified 
according to 
the total spin and orbital angular
momentum of the quarks in the baryon rest frame, which separates the basis into
$s$, $p$, and $d$ waves \cite{Eichmann:2009qa,Eichmann:2009zx,Sanchis-Alepuz:2011egq}. We follow the notation of \cite{Eichmann:2009zx, Sanchis-Alepuz:2014sca}, in which the letters $S$, $P$, $A$ and $V$ indicate the scalar, pseudoscalar, axial or vector structure of a quark pair, while $X$ denotes combinations with total quark spin $3/2$. For instance, $S_{11}$ and $V_{11}$ are $s$ waves, whereas $S_{13}$, $V_{13}$
and $X_{53}$ are $p$ waves. Each basis 
element carries a positive- or negative-energy projector, indicated by a
superscript $\pm$.

The three terms in the sum of \1eq{eq:faddeev} differ solely in the choice of
the spectator quark. Since the color component of the amplitude is antisymmetric,
the Pauli principle requires its remaining spin, momentum, and flavour
structure to be symmetric under the exchange of any two valence quarks. The three
terms are therefore related by a permutation of the quark labels, so that
only one of them has to be computed explicitly, while the other two follow
via simple relabeling \cite{Eichmann:2009qa,Sanchis-Alepuz:2014sca}.

The flavour-dependent kernel preserves this property. A permutation of the
quark labels requires the kernel to be symmetric under the exchange of the two
quarks it connects, which is guaranteed by the invariance of $\Int_{ff'}$ 
under $f \leftrightarrow f'$.
Furthermore, such a permutation only changes the momentum variables
by fixed numerical coefficients. Since $A_f(q^2)$ depends solely on the gluon
momentum, it transforms under this change exactly as $\tilde{\alpha}_{\s T}(q^2)$. The only
modification concerns the flavour content: the integral is evaluated
separately for each quark pair present within the baryon, namely $\ell\ell$,
$\ell s$ or $ss$ with $\ell=u,d$, with each individual term of \1eq{eq:faddeev}
constructed from the quark pair it contains. The same treatment is  required within the traditional RL
framework
whenever the quark masses differ, because the propagators circulating inside the loop
depend directly on the active pair \cite{Sanchis-Alepuz:2014sca}. Crucially, the factor $A_f(q^2)A_{f'}(q^2)$ enters only as a
multiplicative weight in the integrand,
making the computational cost entirely comparable to that of the standard 
RL approximation.

We close this section with a brief comment on the $\Lambda$ and $\Sigma$ states. Both baryons share an identical 
$uds$ valence-quark content, differing 
only in the isospin configuration of the light quark pair, 
with isospin $0$ in the $\Lambda$ and $1$ in the $\Sigma$. The interaction of 
\1eq{eq:kernel} preserves the flavour of each individual quark line, and, because 
$A_u=A_d$ in the isospin-symmetric limit employed here, the resulting kernel treats the $u$ and $d$
quarks identically. Consequently, the method cannot resolve these two distinct 
isospin combinations, causing the $\Lambda$ and $\Sigma$ masses to emerge exactly 
degenerate.
An identical degeneracy is manifested within the traditional RL 
approximation~\cite{Sanchis-Alepuz:2014sca,Eichmann:2016yit}. 
In more comprehensive three-body 
calculations, this degeneracy is lifted by incorporating explicit 
meson-exchange contributions 
between the 
valence quarks~\cite{Fischer:2008sp,Sanchis-Alepuz:2014wea}.

\section{Results}
\label{sec:results}

\subsection{Numerical inputs} \label{sec:inputs}

All dynamical equations are renormalized within the Taylor scheme, 
employing \mbox{$\mu = 4.3$ GeV} as the renormalization scale. 
As previously stated, we work in
the isospin-symmetric limit, where the up and down quarks have identical 
current masses ($m_u = m_d$).
Specifically, the current quark masses are strictly fixed to 
the values tuned within the mesonic domain in~\cite{Gao:2024gdj}:
\begin{equation}
  m_{u/d} = 0.005\ \text{GeV} \,, \qquad m_s = 0.094\ \text{GeV} \,.
  \label{eq:masses}
\end{equation}
These parameters were determined by requiring that the solutions of \eq{eq:gap}, 
when incorporated into the mesonic BSE, successfully 
reproduce the physical pion and kaon masses, as discussed below.

For the modified Taylor coupling we adopt the parametrization established in
\cite{Gao:2024gdj} {\it without} further adjustment. 
Crucially, 
$\tilde{\alpha}_{\s T}(q^2)$ is an RGI 
quantity, making it completely independent of the specific choice of 
renormalization scale $\mu$.
The functional form displayed in
\fig{fig:alphaT} corresponds to the expression 
\be
{\tilde\alpha}_{\s T}(q^2)
= \frac{a_0 q^2 + a_1 q^4 \ln\left(1 + \frac{\Lambda_0^2}{q^2}\right) + a_2 q^4}{1 + a_3 q^2 + a_4 q^4 + a_5 q^6}
+ \frac{4 \pi q^6}{\beta_0 \left(\Lambda_0^6 + q^6 \ln \frac{q^2}{\Lambda_T^2}\right)},
\label{eq:dhat_param}
\ee
with the fit coefficients fixed to $a_0 = 10.35$ GeV$^{-2}$, $a_1=23.69$ GeV$^{-4}$, $a_2=27.94$ GeV$^{-4}$,
$a_3=10.72$ GeV$^{-2}$, $a_4=-2.5$ GeV$^{-4}$, $a_5=29.02$ GeV$^{-6}$, $\Lambda_T=0.5$ GeV, $\Lambda_0=1$ GeV, and 
\mbox{$\beta_0=11-2n_f/3$}
($n_f=4$). 
The 
uncertainties of these fit parameters, originating from the statistical error 
bars of the lattice data for $\alpha_{\s T}(q^2)$ \cite{Zafeiropoulos:2019flq}, dictate the propagation of error 
into our computed baryonic observables, as detailed in \sect{sec:error}.

%
%%%%%%%%%%%%%%%%%%%%%%%%%%%%%%%%%%%%%%%%%%%%%%%%%%%%
\begin{table}[t]
  \centering
  \renewcommand{\arraystretch}{1.45}
  \begin{tabular}{lcccc}
    \hline\hline
     & $m_\pi$ & $m_K$ & $f_\pi$ & $f_K$ \\
    \hline
    This work  & $0.139(2)^{\dagger}$ & $0.495(2)^{\dagger}$
               & $0.130(1)^{\dagger}$ & $0.158(2)$ \\
    Experiment & $0.139$ & $0.494$ & $0.130$ & $0.155$ \\
    \hline\hline
  \end{tabular}
  \caption{Masses and decay constants of the pion and the kaon, in GeV,
  obtained from \1eq{eq:bse} with the inputs of \sect{sec:inputs}. The
  entries marked with a dagger are those employed to fix the current quark
  masses and the normalization of the amplitudes. The experimental values are taken
  from \cite{ParticleDataGroup:2026mpi}.}
  \label{tab:mesons}
\end{table}
%%%%%%%%%%%%%%%%%%%%%%%%%%%%%%%%%%%%%%%%%%%%%%%%%%%%

%%%%%%%%%%%%%%%%%%%%%%%%%%%%%%%%%%%%%%%%%%%%
\begin{table*}[t!]
  \centering
  \renewcommand{\arraystretch}{1.35}
  \setlength{\tabcolsep}{9pt}
  \begin{tabular}{l cccc c cccc}
    \hline\hline
     & \multicolumn{4}{c}{$J^P = \tfrac{1}{2}^{+}$} & &
       \multicolumn{4}{c}{$J^P = \tfrac{3}{2}^{+}$} \\
    \cline{2-5} \cline{7-10}
     & $N$ & $\Lambda$ & $\Sigma$ & $\Xi$ & & $\Delta$ & $\Sigma^{*}$ & $\Xi^{*}$ & $\Omega$ \\
    \hline
    This work  & $0.936(5)$ & $1.097(6)$ & $1.097(6)$ & $1.288(7)$ & & $1.214(7)$ & $1.354(8)$ & $1.510(9)$ & $1.684(12)$ \\
    RL         & $0.930(3)$ & $1.073(1)$ & $1.073(1)$ & $1.235(5)$ & & $1.21(2)$  & $1.33(2)$  & $1.47(3)$ & $1.65(4)$   \\
    Experiment & $0.939$    & $1.116$    & $1.189$    & $1.315$    & & $1.232$    & $1.385$    & $1.533$   & $1.672$     \\
    \hline\hline
  \end{tabular}
  \caption{Masses of the positive-parity octet and decuplet, in GeV, compared
  with the RL results of \cite{Sanchis-Alepuz:2014sca} and with
  experiment \cite{ParticleDataGroup:2026mpi}.}
  \label{tab:spectrum}
\end{table*}
%%%%%%%%%%%%%%%%%%%%%%%%%%%%%%%%%%%%%%%%%%%%%%%%%%%%
Both the mesonic BSEs and the three-body Faddeev equation, \1eq{eq:bse} and \1eq{eq:faddeev}, respectively,  are solved as eigenvalue problems.
The
physical solutions correspond to the on-shell values $P^2 = -M^2$, where the
leading eigenvalue equals unity. With these inputs, the solution of
\1eq{eq:bse} in the pseudoscalar channel returns the masses and decay
constants collected in \tab{tab:mesons}. The two meson masses and the normalization
of $f_\pi$ are precisely the quantities used to fix the current quark masses \1eq{eq:masses}. Therefore,
the only prediction of this table is the value of $f_K$, which reproduces its
experimental value at the level of two percent. Thus, the quoted numbers completely determine the interaction \1eq{eq:kernel} that enters in the baryon calculation.

With the exception of the $\Xi^{*}$, the complex momenta probed by
\1eq{eq:faddeev} lie within the domain of analyticity of $S_{\!f}$; therefore the corresponding masses
are obtained directly on shell. In the case of the $\Xi^{*}$, the domain probed by the
light quark reaches the complex conjugate poles of $S_{\!f}$ before the
physical mass is attained; thus, the mass of $\Xi^{*}$ is obtained 
from a set of smaller $P^2$ values 
by extrapolating the leading
eigenvalue to unity. 

All integrals are evaluated numerically, with $30$ points for each of the
momentum variables, while $8$ points are employed for each of
the angular variables. The angular dependence of the dressing functions in
\1eq{eq:basis} is expanded in Chebyshev polynomials of the second kind.  It turns out that retaining the first three of them for each of the three angles $\{z\}$ is sufficient to reach convergence in the calculation of baryon masses.

\subsection{Numerical results} \label{sec:numres}

\begin{figure}[t]
  \centering
  \includegraphics[width=\columnwidth]{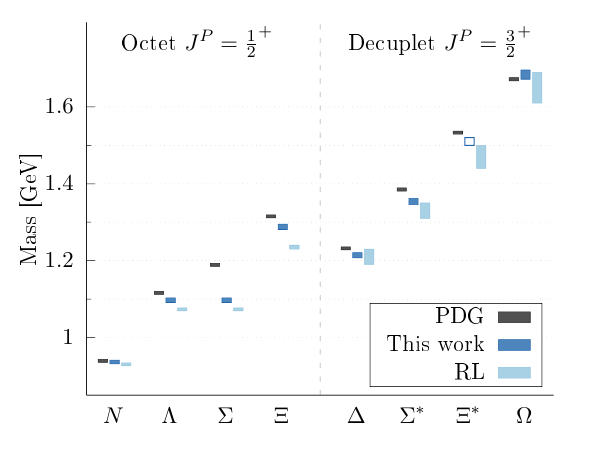}
  \caption{Masses of the positive-parity octet and decuplet obtained in this
  work, compared with the RL results of \cite{Sanchis-Alepuz:2014sca} and with
  experiment \cite{ParticleDataGroup:2026mpi}. The height of each box indicates
  the uncertainty (see Sec.~\ref{sec:error}). The open box
  marks the $\Xi^{*}$, whose mass is obtained by extrapolation of the
  eigenvalue curve.}
  \label{fig:spectrum}
\end{figure}

With the ingredients of \sect{sec:inputs} in place, we solve \1eq{eq:faddeev}
for the octet and decuplet states; the resulting baryon masses are collected in
\tab{tab:spectrum} and displayed in \fig{fig:spectrum}. Since in 
the isospin-symmetric limit the members of
each isospin multiplet are degenerate, a single representative for each is quoted. In the above table 
and figure we also report 
the corresponding masses obtained within the RL approximation \cite{Sanchis-Alepuz:2014sca},  together with the experimental values taken from 
\cite{ParticleDataGroup:2026mpi}. 

%%%%%%%%%%%%%%%%%%%%%%%%%%%%%%%%%%%%%%%%%%%%%%%%%%%%
%%%%%%%%%%%%%%%%%%%%%%%%%%%%%%%%%%%%%%%%%%%%%%%%%%%%

As can be seen in \fig{fig:spectrum}, the computed masses agree with
experiment at the level of a few percent, with the exception of the $\Sigma$. With respect to the RL, the masses of the nucleon and the $\Delta$ change by only a few MeV, while those of the states containing strange quarks shift by a few tens of MeV. Thus, the differences between the two calculations arise mainly from
the strange sector.

The splitting between the nucleon and the $\Delta$ is essentially the same as
in the RL approximation ($0.278$~GeV against $0.280$~GeV), and somewhat below the experimental value of 
$0.293$~GeV. This is consistent with the fact that both states contain only
light quarks, so that the 
flavour-dependence of the kernel 
is controlled by 
$\Int_{\!\ell\ell}$ alone.

The $\Lambda$ and the $\Sigma$ baryons emerge  degenerate, as anticipated in
\sect{sec:faddeev}, sharing a mass that lies below both measured values, thus giving rise to a larger deviation
for the $\Sigma$. To properly account 
for the observed mass splitting, 
the flavor-diagonal interaction kernel employed here must be 
supplemented with suitable flavour non-diagonal contributions \cite{Sanchis-Alepuz:2014sca}, which is beyond the scope of the present study.

\begin{figure}[t!]
  \centering
\includegraphics[width=1.0\columnwidth]{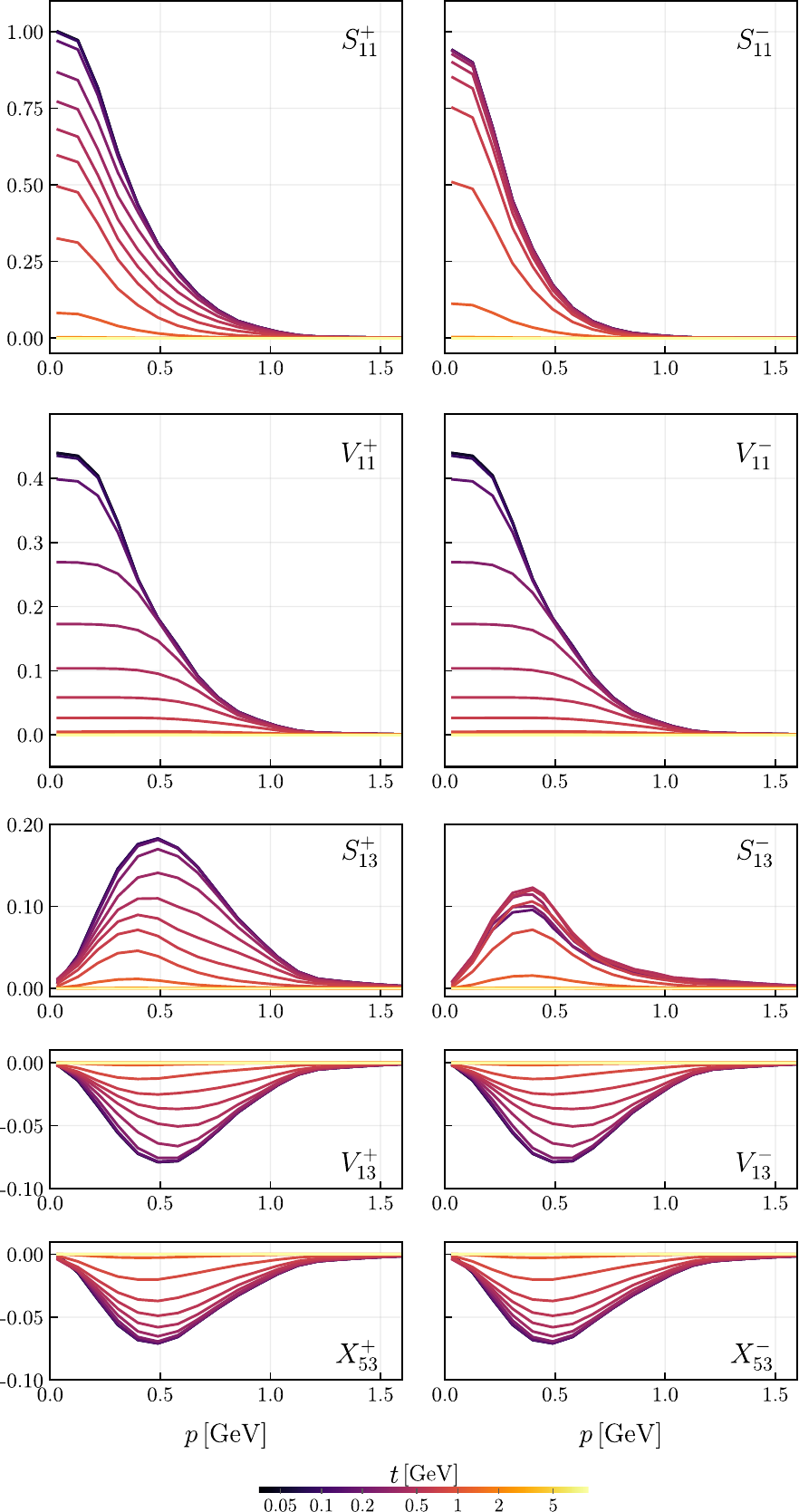}
\caption{Dressing functions of the dominant covariants in the
mixed-antisymmetric component of the nucleon Faddeev amplitude,
\1eq{eq:basis}, at zeroth order in the Chebyshev expansion. Each curve
corresponds to a fixed value of the relative momentum $t$, indicated by the
colour scale.}
  \label{fig:nucleon}
\end{figure}

In \fig{fig:nucleon} we display 
the dominant dressing functions contributing to the  mixed-antisymmetric component $\Psi_{\rm MA}$ of the nucleon Faddeev amplitude, see \1eq{eq:basis}.
The two $s$ waves, $S_{11}$ and $V_{11}$,
dominate this amplitude, while the $p$ waves $S_{13}$, $V_{13}$, and $X_{53}$
are considerably smaller and peak at intermediate momenta, providing a
sizeable $p$-wave admixture to the nucleon amplitude. In agreement with
previous studies \cite{Eichmann:2009qa,Eichmann:2016yit}, the positive- and
negative-energy partners come out comparable in size. The negative-energy components have no counterpart in a non-relativistic
description, so their size reflects the relativistic character of the
three-quark system. The mixed-symmetric component 
$\Psi_{\rm MS}$ displays a similar qualitative behavior, and will not be shown here.  

\begin{figure*}[t!]
  \centering
\includegraphics[scale=1]{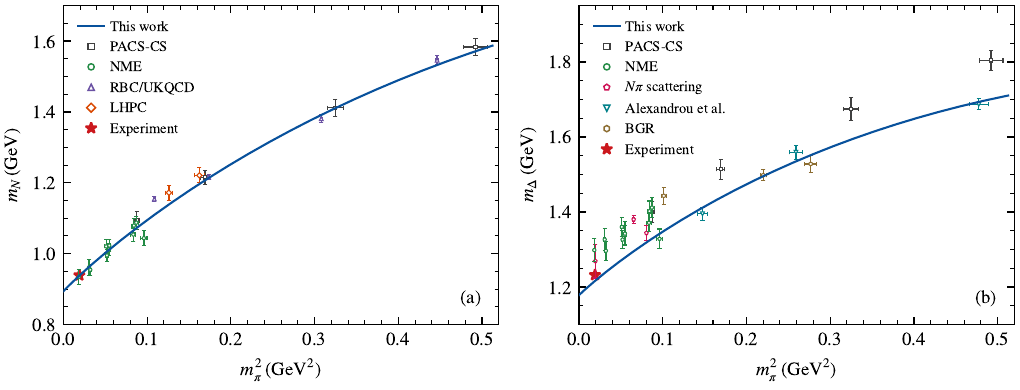}
{\caption{Left: the mass of the nucleon as a function of $m_\pi^2$. Right: the
mass of the $\Delta$ as a function of $m_\pi^2$. In both panels, the solid
lines show the results of this work, and the symbols the lattice QCD results of
\cite{PACS-CS:2008bkb,Yoo:2026uul,Yamazaki:2009zq,Syritsyn:2009mx,Alexandrou:2009hs,Gattringer:2008vj,Andersen:2017una,Silvi:2021uya,Alexandrou:2023elk}.
The stars mark the experimental values \cite{ParticleDataGroup:2026mpi}.}
\label{fig:masses_mpi2}}
\end{figure*}

The dependence of the nucleon and $\Delta$ masses on the light-quark mass
provides an additional test of the effective interaction. In particular, we vary the current mass of the
light quarks while keeping the rest of the input fixed and obtain the
corresponding pion mass from \1eq{eq:bse}. The resulting curves are shown in
\fig{fig:masses_mpi2} as functions of $m_\pi^2$, together with the corresponding lattice QCD results \cite{PACS-CS:2008bkb,Yoo:2026uul,Yamazaki:2009zq,Syritsyn:2009mx,Alexandrou:2009hs,Gattringer:2008vj,Andersen:2017una,Silvi:2021uya,Alexandrou:2023elk}. 
The nucleon-mass curve follows the lattice data over the entire range of pion masses
considered, up to $m_\pi \approx 0.7$~GeV. In the chiral limit it decreases to
$0.900$~GeV, about $36$~MeV below its value at the physical point. The $\Delta$-mass curve shows a similar behaviour, lying 
$18$~MeV below the experimental value
at the physical point. For larger pion masses it agrees with the lattice results of
\cite{Alexandrou:2009hs,Gattringer:2008vj}, and lies below those of
\cite{PACS-CS:2008bkb}.

We conclude with a brief remark on the limitations arising from retaining 
solely the classical tensor structure of the quark-gluon vertex, a feature 
that the present scheme shares with the RL approximation.

 In the
language of the quark model, the ground states of the octet and the decuplet
are dominated by $s$ waves. For the nucleon and the $\Delta$, RL-type kernels have been found to be reliable precisely in these channels
\cite{Eichmann:2016hgl}.
The situation changes markedly in other channels. In the meson sector, the axial-vector states, which in the quark model are $p$ waves, deviate appreciably from experiment. The same pattern
appears for baryons, where the parity partner of the nucleon is underestimated by roughly twenty percent. In addition, the level ordering between the Roper
and the $N(1535)$ is inverted \cite{Eichmann:2016hgl}; this artifact has been attributed to the vector-vector character of the effective gluon exchange, which does not properly capture the spin-dependent
parts of the interaction, a limitation that 
carries over directly to our framework.
Likewise, the ordering of the radial excitations of
the pion and the kaon is restored once the full tensor structure of the
quark-gluon vertex is included \cite{Ferreira:2026gbe}. Observables of this kind are sensitive to the full tensorial structure of the quark-gluon interaction, and lie beyond the scope of the present work.

%%%%%%%%%%%%%%%%%%%%%%%%%%%%%%%%%%%%%%%%%%%%%%%%%%%%
\subsection{Error analysis}
\label{sec:error}

As indicated by the error
bars in \fig{fig:alphaT},
the lattice uncertainties associated with the Taylor coupling
$\alpha_{\s T}(q^2)$ are transferred to $\tilde\alpha_{\s T}(q^2)$.
These error bars, in turn, determine the uncertainty of the parameters of
\1eq{eq:dhat_param}, which are fitted to the lattice points. 
The blue band 
indicated in \fig{fig:alphaT} is obtained by 
varying the fit 
parameters by one standard deviation around their central values; the resulting width of the band is compatible with the lattice error bars. 
To estimate the errors of our approach, we
repeat the entire  calculation,
substituting for $\tilde\alpha_{\s T}(q^2)$
in the dynamical equations 
(gap equation, meson BSE, and Faddeev equation) the upper and lower limits of this band, keeping the current quark masses
fixed. The resulting spread of the hadron masses is the uncertainty quoted in
\tab{tab:mesons}, \tab{tab:spectrum} and \fig{fig:spectrum}; in all cases, it is less than 1\%.

As discussed in \sect{sec:inputs}, in the case of the $\Xi^{*}$, the complex conjugate poles of $S_{\!f}$ limit the range of 
$M$ accessible to the Faddeev equation. The leading eigenvalue is therefore computed for smaller values of $M$ and
extrapolated to unity by means of the ``Schlessinger point method'' 
\cite{Schlessinger:1968vsk,Tripolt:2018xeo}. We estimate the uncertainty of
this extrapolation from the spread of the masses obtained when different
subsets of the calculated eigenvalues are used. This uncertainty is shown as
an open box in \fig{fig:spectrum}, 
and is of the order of 
1\%.

\section{Conclusions} \label{sec:outlook}

In this work we have employed the flavour-dependent interaction of
\cite{Gao:2024gdj} in the baryon sector, 
where it was used 
in the three-body Faddeev equation to determine 
the masses of the
positive-parity octet and decuplet. This particular 
interaction strength 
is built out of a modified Taylor effective charge 
multiplied by 
quark dressing functions that encode
the flavor content of the two interacting valence quarks.

The resulting spectrum agrees with experiment at the level of a few percent,
with the exception of the $\Sigma$, whose mass lies around 8\% below the measured value. 
Compared to the RL treatment, the masses of the 
non-strange states are barely affected, whereas those of states carrying strange 
valence quarks shift systematically upward by a few tens of MeV, moving closer to the 
experimental values.
 The $\Delta$-$N$ splitting is
essentially unchanged, and the $\Lambda$ and the $\Sigma$ remain degenerate,
as dictated by the flavour-diagonal structure of the kernel employed.

The present kernel retains only the classical tensor structure of the
quark-gluon vertex. Kernels of this type have proved reliable in the channels
of the nucleon and the $\Delta$, but not in other spin-parity channels, such
as those of the negative-parity baryons \cite{Eichmann:2016hgl}. Incorporating 
the full vertex content, as recently executed for mesons within the ``symmetric-vertex approximation''
\cite{Ferreira:2026gbe,Ferreira:2026zpr}, should allow these channels to be reliably addressed. 

Finally, the present framework can be applied to the electromagnetic form factors
of the octet and decuplet, as already done for mesons \cite{Miramontes:2025vzb},
and to baryons containing heavy quarks, where the asymmetry between the
flavours is considerably larger.

\section*{Acknowledgements}
We thank Raul~D.~Torres for useful discussions.
This work is funded by the Spanish MICINN grants PID2020-113334GB-I00 and
PID2023-151418NB-I00,  
the Generalitat Valenciana grant CIPROM/2022/66,
and CEX2023-001292-S by MCIU/AEI.
H. S-A is supported by Silicon Austria Labs (SAL), owned by the Republic of Austria,  the Styrian Business Promotion Agency (SFG), the federal state of Carinthia, the Upper Austrian Research (UAR), and the Austrian Association for the Electric and Electronics Industry (FEEI).

\bibliographystyle{elsarticle-num-names}
\bibliography{bibliography.bib}
\end{document}